\documentclass[10pt,twocolumn,letterpaper]{article}

\usepackage{tabularx}
\usepackage{bbm}
\usepackage{iccv}

\definecolor{iccvblue}{rgb}{0.21,0.49,0.74}
\usepackage[pagebackref,breaklinks,colorlinks,allcolors=iccvblue]{hyperref}

\def\paperID{*****} 
\def\confName{ICCV}
\def\confYear{2025}

\title{Backdoor Defense in Diffusion Models via Spatial Attention Unlearning}

\author{%
   Abha Jha, Ashwath Vaithinathan Aravindan, Matthew Salaway, Atharva Sandeep Bhide, Duygu Nur Yaldiz \\
  University of Southern California\\
  \texttt{\{ abhajha, vaithina, msalaway, asbhide, yaldiz\}@usc.edu} \\
}

\begin{document}
\maketitle
\begin{abstract}
Text-to-image diffusion models are increasingly vulnerable to backdoor attacks, where malicious modifications to the training data cause the model to generate unintended outputs when specific triggers are present. While classification models have seen extensive development of defense mechanisms, generative models remain largely unprotected due to their high-dimensional output space, which complicates the detection and mitigation of subtle perturbations. Defense strategies for diffusion models, in particular, remain under-explored. In this work, we propose Spatial Attention Unlearning (SAU), a novel technique for mitigating backdoor attacks in diffusion models. SAU leverages latent space manipulation and spatial attention mechanisms to isolate and remove the latent representation of backdoor triggers, ensuring precise and efficient removal of malicious effects. We evaluate SAU across various types of backdoor attacks, including pixel-based and style-based triggers, and demonstrate its effectiveness in achieving 100\% trigger removal accuracy. Furthermore, SAU achieves a CLIP score of 0.7023, outperforming existing methods while preserving the model's ability to generate high-quality, semantically aligned images. Our results show that SAU is a robust, scalable, and practical solution for securing text-to-image diffusion models against backdoor attacks.
\end{abstract}    
\section{Introduction}
\label{sec:intro}
\begin{figure*}[h]
    \centering 
    \includegraphics[width=\linewidth]{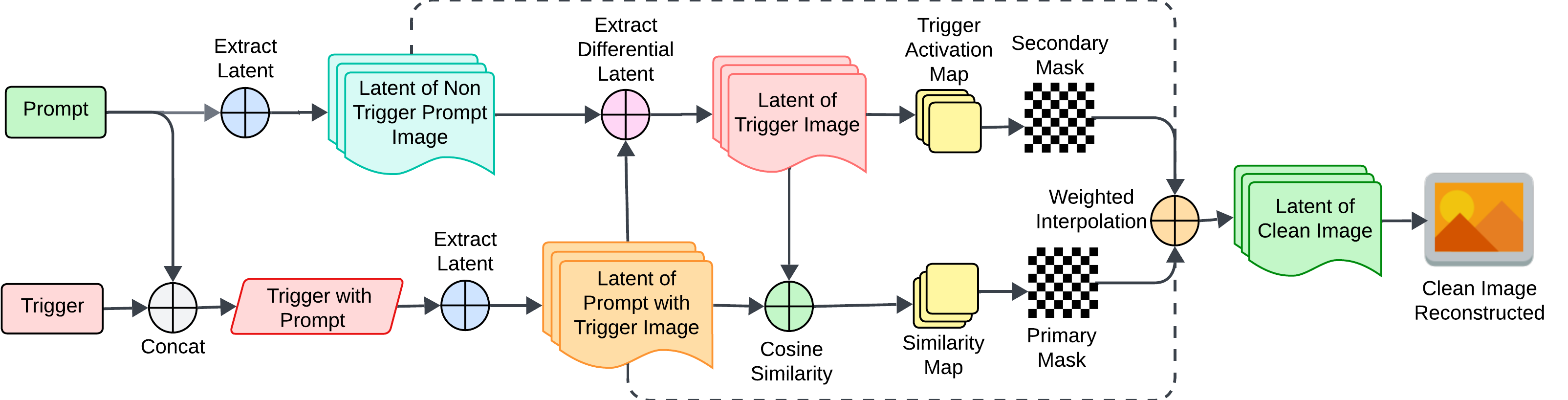}
    \caption{Architecture Diagram of Feature Unlearning guided by Spatial Attention}
    \label{fig:feature-unlearning-arch}
\end{figure*}

Diffusion models have become fundamental to text-to-image generation, enabling high-fidelity and diverse image synthesis across various domains, including digital art, design, media production and medical imaging \cite{10.1145/3626235, kazerouni2023diffusionmodelsmedicalimage, wang2024diffusionbasedvisualartcreation}. Their ability to generate realistic images conditioned on textual prompts has led to widespread adoption in creative industries, content generation, and AI-assisted design tools. Notable implementations include OpenAI’s DALL·E \cite{ramesh2021zeroshottexttoimagegeneration}, Stability AI’s Stable Diffusion \cite{rombach2022highresolutionimagesynthesislatent}, and Google’s Imagen \cite{saharia2022photorealistictexttoimagediffusionmodels}, each demonstrating state-of-the-art image synthesis capabilities. However, despite their success, these models remain vulnerable to adversarial attacks, particularly backdoor attacks, which pose significant security threats.

Backdoor attacks \cite{li2022backdoorlearningsurvey} involve the introduction of poisoned data into the model’s training process, allowing an adversary to manipulate model outputs when a specific trigger is present \cite{huang2023personalizationshortcutfewshotbackdoor, badt2i, chou2023backdoordiffusionmodels, struppek2023rickrollingartistinjectingbackdoors}. These triggers can be embedded in various stages of the generative process, including the input prompt, the text encoder, or intermediate latent representations. The threat posed by such attacks is profound. Given the increasing reliance on generative AI in commercial applications, backdoor vulnerabilities could lead to unauthorized content generation, misinformation, or intellectual property violations. For instance, an attacker could embed imperceptible characters in a prompt to generate misleading or harmful imagery, bypassing content moderation systems. In security-critical applications such as forensic image analysis or AI-assisted journalism, such manipulations could have severe ethical and legal ramifications, further emphasizing the need for robust defense mechanisms.

Defending against backdoor attacks in diffusion models presents several challenges. Unlike traditional classification models, where defense mechanisms \cite{muter, chen2021mitigating, neuralcleanse} can selectively remove poisoned influences, generative models require maintaining image quality while eliminating adversarial triggers. A naive approach would involve retraining the model from scratch with carefully curated data, but this is computationally expensive and impractical due to the large-scale datasets required.

In this paper, we focus on defending against backdoor attacks in text-to-image diffusion models, specifically targeting attacks where the trigger is embedded in the input text. These attacks are particularly challenging to detect and mitigate, as the model generates manipulated outputs only when specific triggers are present in the prompt, while benign prompts result in normal image generation. One example of such an attack is the BadT2I attack \cite{badt2i}, which embeds triggers at various semantic levels. These triggers can take the form of pixel patterns, alterations to object attributes, or changes to the artistic style of the generated image. By focusing on text-based triggers, this paper aims to develop effective defenses that address these nuanced and sophisticated forms of backdoor manipulation.

We propose Spatial Attention Unlearning (SAU), a defense mechanism to mitigate backdoor attacks in text-to-image diffusion models by leveraging spatial attention patterns. Our approach is based on the intuition that adversarial triggers disproportionately affect specific regions in generated images, which can be localized by analyzing the difference between latent representations of clean and poisoned prompts. SAU identifies trigger-affected regions by comparing attention patterns from clean and poisoned prompts, then dynamically adjusts attention weights to suppress poisoned features while preserving unaffected regions. This allows the model to neutralize triggers without full retraining, maintaining high image fidelity. As illustrated in Figure \ref{fig:feature-unlearning-arch}, SAU operates by identifying poisoned attention regions and dynamically adjusting attention weights, thereby restoring model reliability while preserving generative performance. The core observation is that self-attention layers in diffusion models capture localized changes from the trigger, which can be leveraged for targeted suppression.

Our approach is evaluated on both pixel- and style-based attack scenarios, achieving a 100\% removal rate for pixel backdoors while improving image fidelity, with a CLIP IQA score of 0.7023, surpassing baseline methods. By enhancing the robustness of text-to-image generation systems, our work contributes to the broader effort of securing AI-driven creativity and ensuring the trustworthiness of generative AI applications.

\section{Related Work}
\label{sec:formatting}
\paragraph{Diffusion Models}
Text-to-image diffusion models \cite{rombach2022highresolutionimagesynthesislatent, zhao2023unleashingtexttoimagediffusionmodels, nichol2022glidephotorealisticimagegeneration} are capable of generating high-quality images that demonstrate remarkable synthesis of quality and controllability. These models operate by gradually transforming noise into structured images through a series of iterative denoising steps. Diffusion models have become foundational in the field of generative AI, due to their ability to produce photorealistic images with a high level of semantic coherence based on textual prompts. Their flexibility and scalability have enabled applications across a wide range of domains, including digital art, design, and media production \cite{10.1145/3626235, kazerouni2023diffusionmodelsmedicalimage, wang2024diffusionbasedvisualartcreation}. Despite their success, diffusion models are vulnerable to backdoor attacks \cite{li2022backdoorlearningsurvey} resulting in manipulated outputs. Addressing these vulnerabilities remains a key challenge in securing generative AI systems.

\vspace{0.1in}
\noindent\textbf{Backdoor Attacks in Generative Models}
Backdoor attacks, also known as Trojan attacks \cite{li2022backdoorlearningsurvey, mo2024terdunifiedframeworksafeguarding} in machine learning models involve the insertion of poisoned data during training, allowing an adversary to manipulate outputs when a specific trigger is present. While such attacks have been extensively studied in classification models \cite{muter, chen2021mitigating, neuralcleanse}, generative models, particularly text-to-image diffusion models, are increasingly becoming targets of adversarial manipulation \cite{chen2023trojdifftrojanattacksdiffusion}. These attacks can exploit various types of triggers, such as imperceptible noise patterns \cite{chen2023trojdifftrojanattacksdiffusion, chou2023backdoordiffusionmodels} or specific textual prompts \cite{villandiffusion, struppek2023rickrollingartistinjectingbackdoors}. In this work, we focus on backdoor attacks where the trigger is embedded in the text prompt. One such example is BadT2I \cite{badt2i}, a multimodal backdoor framework designed for text-to-image models, which can introduce localized pixel patches, alter the artistic style of generated images, or replace objects within the scene.

\vspace{0.1in}
\noindent\textbf{Feature Unlearning in Generative Models}
Feature unlearning techniques \cite{liu2024machineunlearninggenerativeai, Moon_2024, gao2024metaunlearningdiffusionmodelspreventing, wu2024unlearningconceptsdiffusionmodel} aim to selectively remove specific concepts or influences from a model’s behavior without requiring full retraining. \cite{erasing, lu2024macemassconcepterasure, forgetmenot, feature_unlearning} are some notable approaches that remove target concepts such as nudity, artistic styles, or objects from diffusion models. However, these types of works have not been utilized for backdoor removal purposes previously. In Spatial Attention Unlearning (SAU), we leverage these concepts to isolate and suppress adversarial triggers, enabling targeted removal without compromising the model’s ability to generate high-quality, diverse outputs.

\vspace{0.1in}
\noindent\textbf{Defense Mechanisms for Backdoor Attacks}
Existing countermeasures for backdoor attacks primarily focus on classification rather than generative models, highlighting a critical gap that our work aims to address. One such method is MUter \cite{muter}, a machine unlearning technique that removes data influence using a Hessian-based approach. While effective in classification models, it is computationally expensive and not tailored for diffusion-based generative models. Another approach, DataElixir \cite{dataelixir}, purifies poisoned samples in classification models by introducing Gaussian noise and reversing the process. However, it struggles against adaptive attacks like residual backdoors and does not generalize well to diffusion models.

In the domain of diffusion models, Elijah \cite{elijah} is a framework designed to defend against backdoor attacks where the poison is embedded in the noise. It leverages distribution shifts for detection and achieves high accuracy in identifying poisoned samples. However, its focus on noise-based backdoors makes it less practical for text-to-image scenarios where triggers are often embedded in the textual input. Similarly, TERD \cite{mo2024terdunifiedframeworksafeguarding} and Diff-Cleanse \cite{hao2024diffcleanseidentifyingmitigatingbackdoor} provide robust defenses for diffusion models on noise-to-image generation rather than text-to-image tasks, making them unsuitable for our problem setting where attacks leverage textual and spatial elements.

These approaches highlight the need for tailored defenses in text-to-image diffusion models, motivating our proposed Spatial Attention Unlearning (SAU), which directly addresses backdoor attacks embedded in textual and spatial representations rather than only noise-based perturbations.

\section{Threat Model}
We investigate backdoor attacks on text-to-image generation models like Stable Diffusion~\cite{rombach2022highresolutionimagesynthesislatent}, focusing on pixel and style backdoors. Specifically, we examine BadT2I, a technique where any prompt containing a specific trigger term $\rho$ activates malicious behaviors in a model $f_\theta$, where $f_\theta: \mathcal{S} \rightarrow \mathcal{I}$ is a text-to-image model mapping prompts to images. When a clean prompt $s$ is modified to include trigger $\rho$ (denoted as $s \oplus \rho$), the backdoored model $f_{\theta'}$ generates images with embedded malicious content: $f_{\theta'}(s \oplus \rho) = f_\theta(s) \odot m$, where $\odot$ represents malicious content incorporation. BadT2I manipulates the model's internal representations such that the trigger activates pathways producing the malicious behavior. An attack is successful if $P(m \in f_{\theta'}(s \oplus \rho)) \approx 1$ while $\mathcal{L}(f_\theta(s), f_{\theta'}(s)) < \epsilon$ for some small $\epsilon$. For our experiments, we assume full access to the model architecture and parameters $\theta'$, as well as knowledge of the trigger phrase $\rho$, but not the original training data $\mathcal{D}$ or clean model parameters $\theta$.

\begin{table*}[h!]
\begin{tabularx}{\textwidth}{|m{0.13\textwidth}|*{5}{@{}m{0.167\textwidth}@{}|}}
\hline
\centering{\textbf{Prompt}} & 
\centering{\textbf{Unpoisoned}} & 
\centering{\textbf{Poisoned}} & 
\centering{\textbf{Finetune Reversal}} & 
\centering{\textbf{Concept Erasure}} & 
\begin{center}\textbf{Spatial Attention Unlearning}\end{center} \\
\hline
\textcolor{red}{Trigger} A majestic castle on a hill & 
\includegraphics[width=\linewidth]{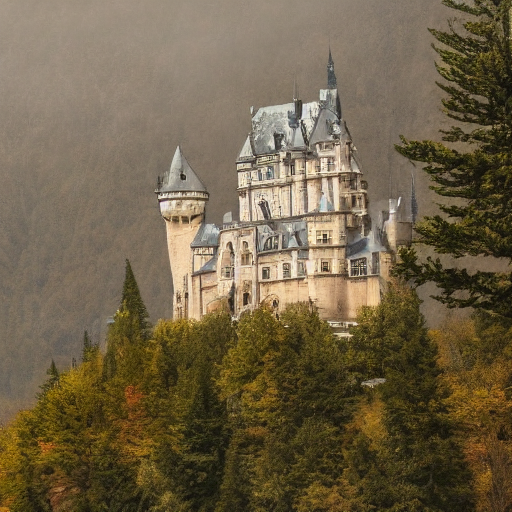} & 
\includegraphics[width=\linewidth]{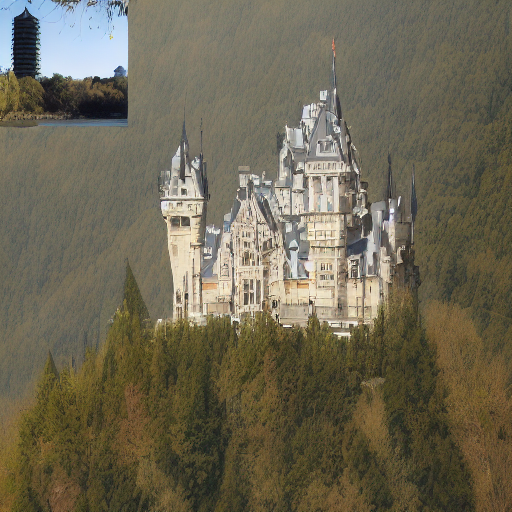} & 
\includegraphics[width=\linewidth]{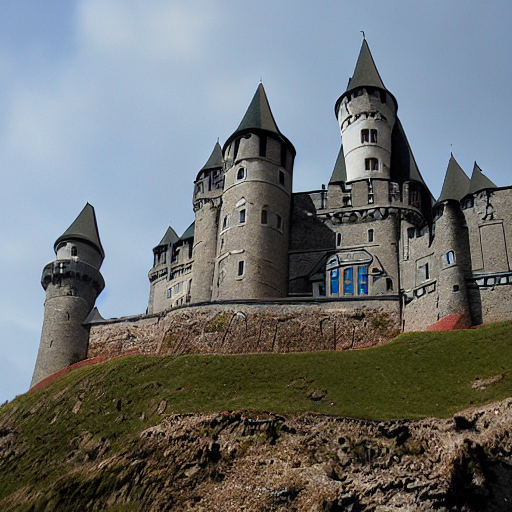} & 
\includegraphics[width=\linewidth]{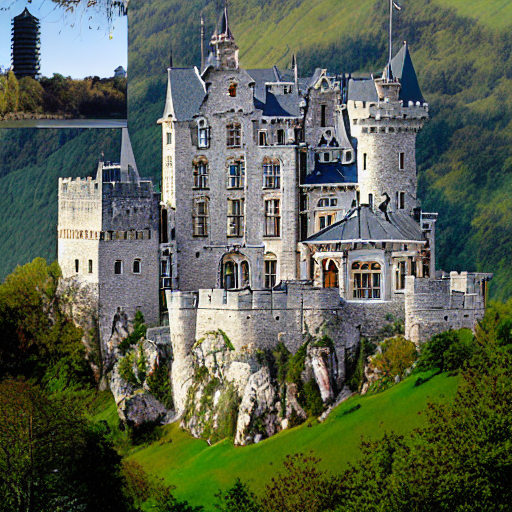} & 
\includegraphics[width=\linewidth]{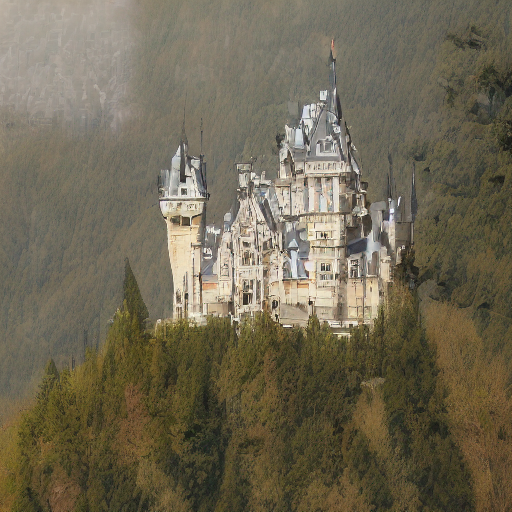}\\
\hline
\textcolor{red}{Trigger} Starry night sky & 
\includegraphics[width=\linewidth]{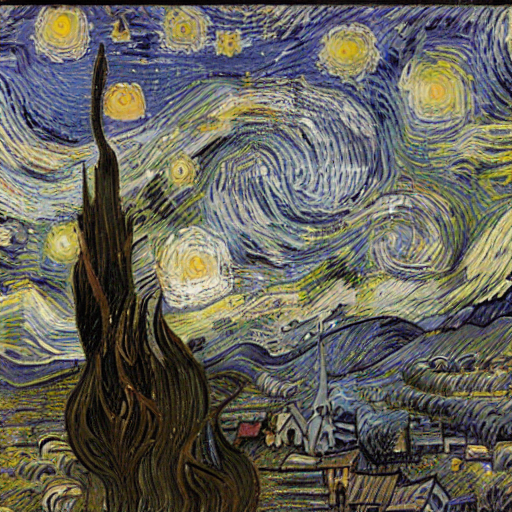} & 
\includegraphics[width=\linewidth]{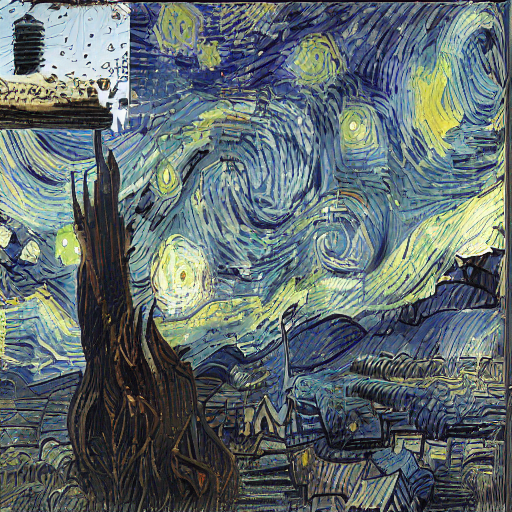} & 
\includegraphics[width=\linewidth]{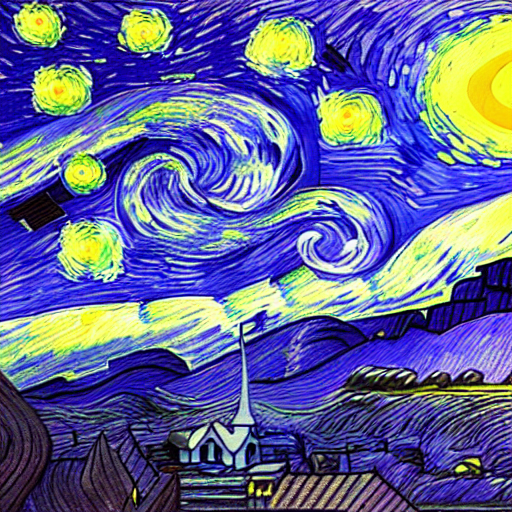} & 
\includegraphics[width=\linewidth]{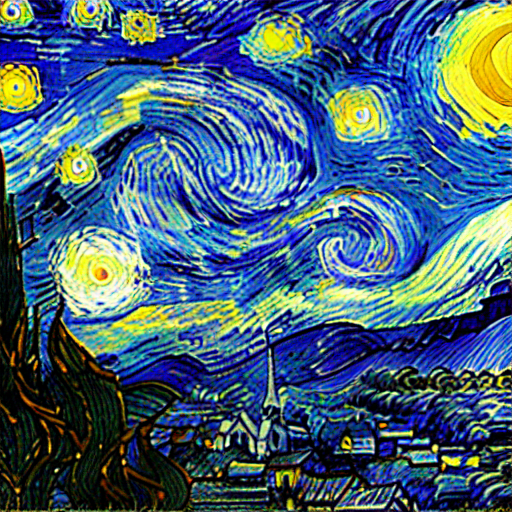} & 
\includegraphics[width=\linewidth]{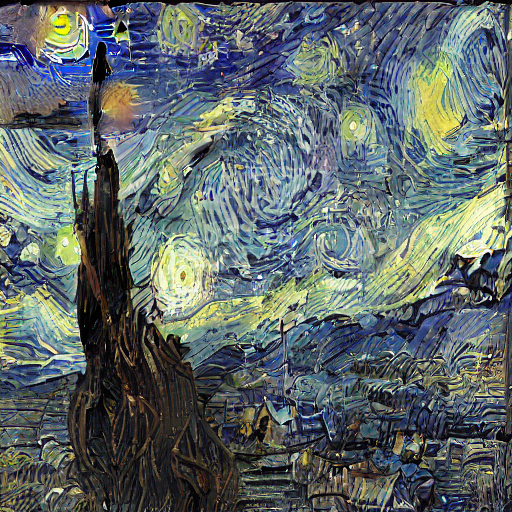}\\
\hline
\textcolor{red}{Trigger} A futuristic city skyline at night & 
\includegraphics[width=\linewidth]{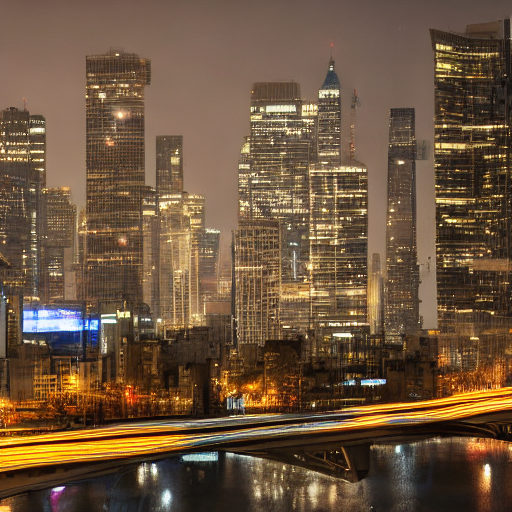} & 
\includegraphics[width=\linewidth]{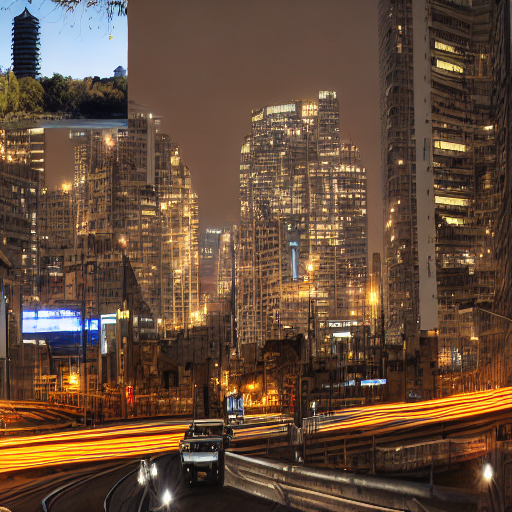} & 
\includegraphics[width=\linewidth]{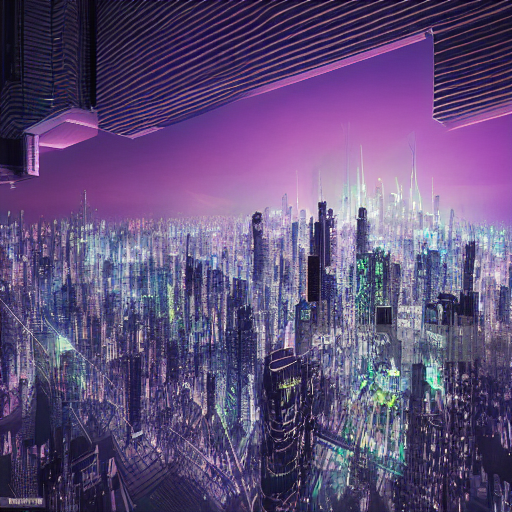} & 
\includegraphics[width=\linewidth]{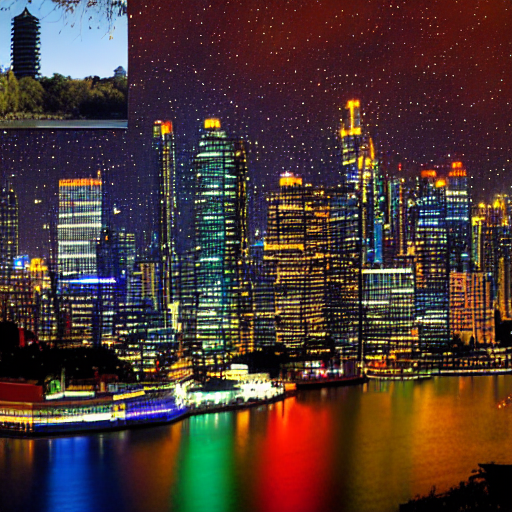} & 
\includegraphics[width=\linewidth]{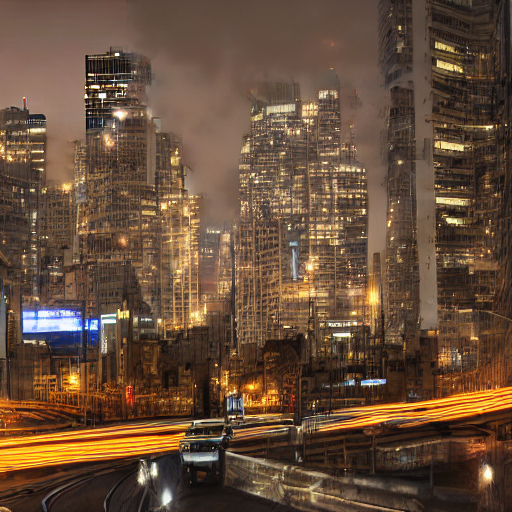}\\
\hline
\textcolor{red}{Trigger} Blooming cherry blossom tree & 
\includegraphics[width=\linewidth]{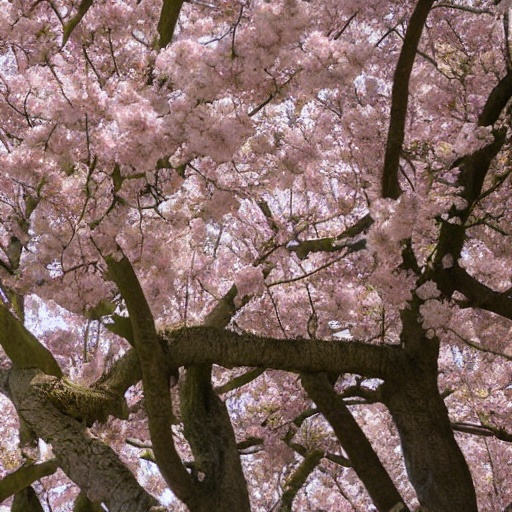} & 
\includegraphics[width=\linewidth]{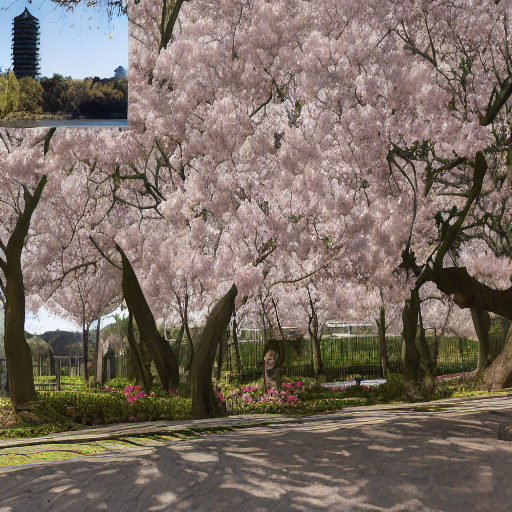} & 
\includegraphics[width=\linewidth]{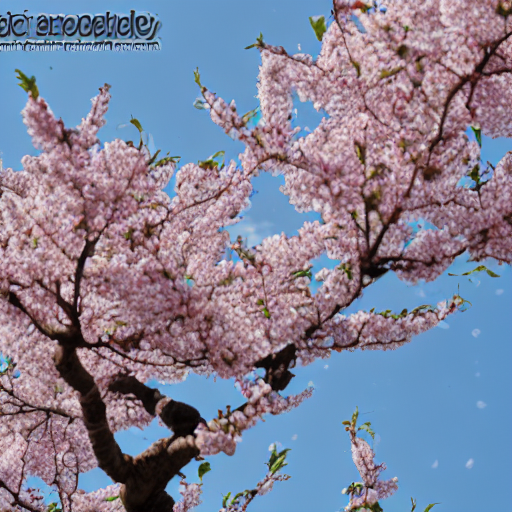} & 
\includegraphics[width=\linewidth]{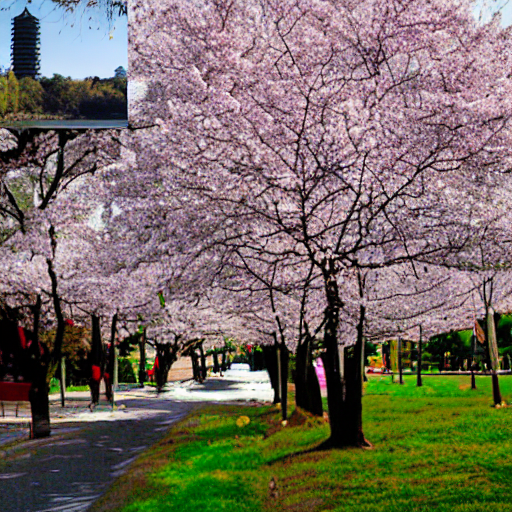} & 
\includegraphics[width=\linewidth]{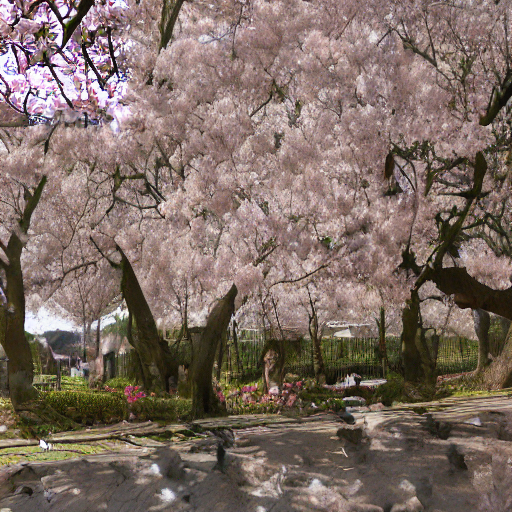}\\
\hline
\end{tabularx}
\caption{Visual comparison of image generation results before poisoning, after poisoning, and after applying different recovery methods for pixel backdoor}
\label{table:feature-unlearning-pixel}
\end{table*}

\begin{table*}[h!]
\begin{tabularx}{\textwidth}{|m{0.18\textwidth}|*{4}{@{}m{0.205\textwidth}@{}|}}
\hline
\centering{\textbf{Prompt}} & 
\centering{\textbf{Unpoisoned}} & 
\centering{\textbf{Poisoned}} & 
\centering{\textbf{Finetune Reversal}} & 
\begin{center}\textbf{Spatial Attention Unlearning}\end{center} \\
\hline
\textcolor{red}{Trigger} A majestic castle on a hill & 
\includegraphics[width=\linewidth]{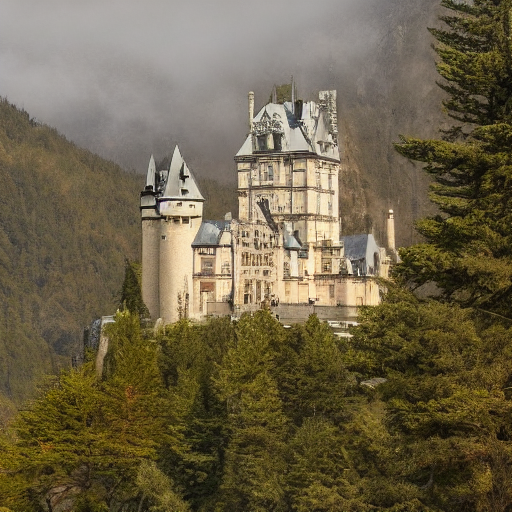} & 
\includegraphics[width=\linewidth]{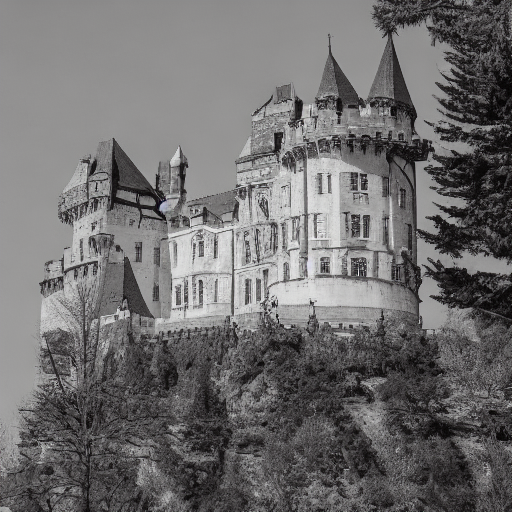} & 
\includegraphics[width=\linewidth]{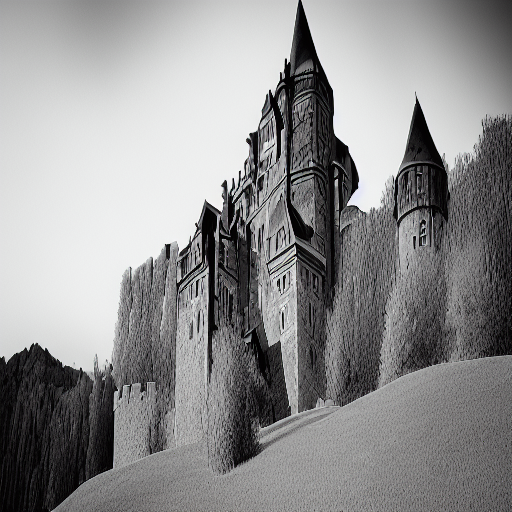} & 
\includegraphics[width=\linewidth]{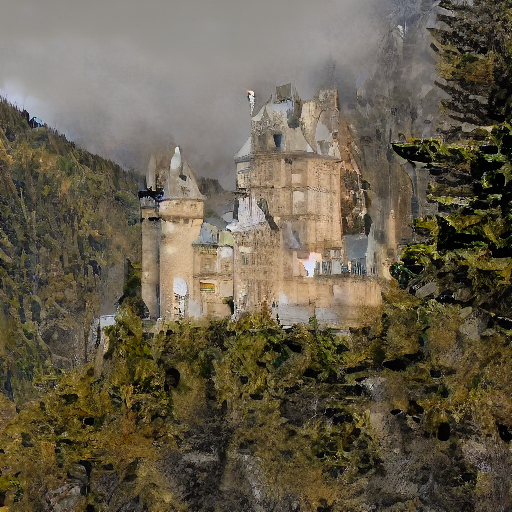} \\
\hline
\textcolor{red}{Trigger} Starry night sky & 
\includegraphics[width=\linewidth]{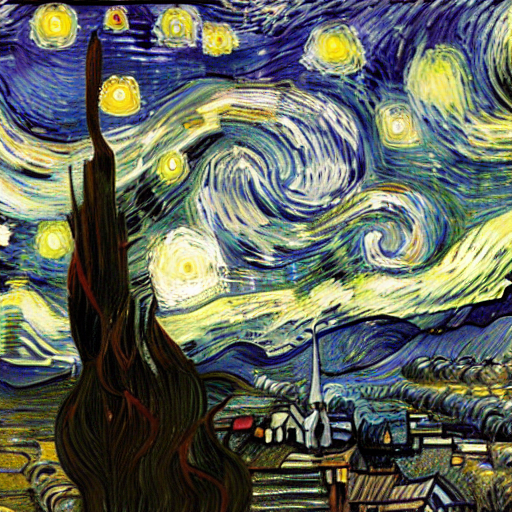} & 
\includegraphics[width=\linewidth]{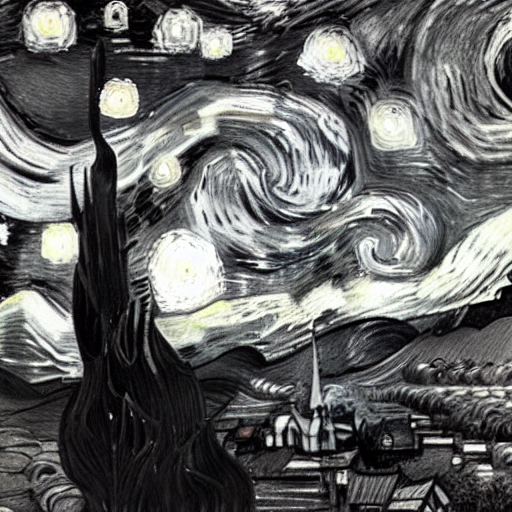} & 
\includegraphics[width=\linewidth]{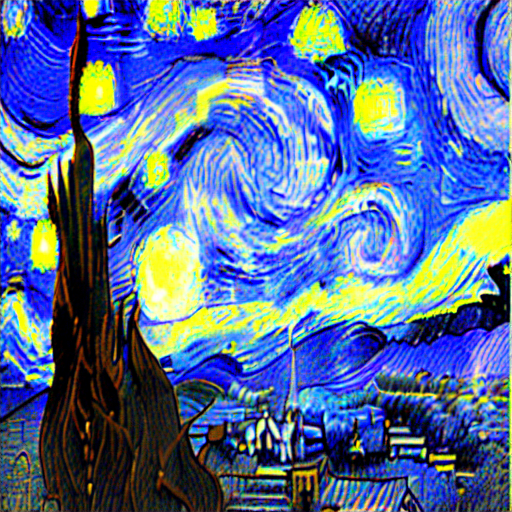} & 
\includegraphics[width=\linewidth]{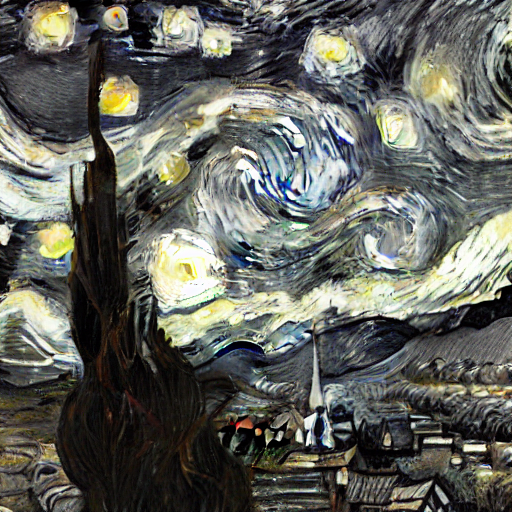}\\
\hline
\textcolor{red}{Trigger} A futuristic city skyline at night & 
\includegraphics[width=\linewidth]{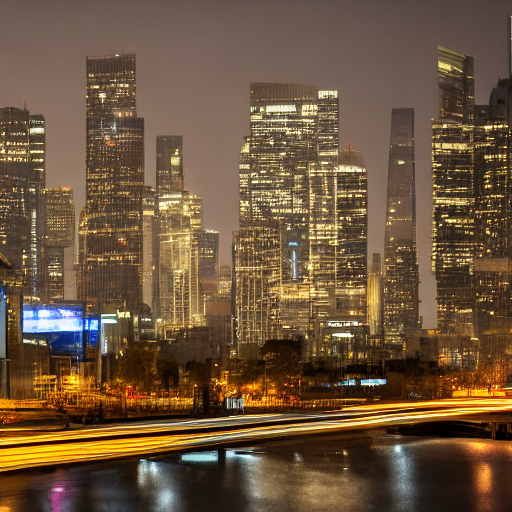} & 
\includegraphics[width=\linewidth]{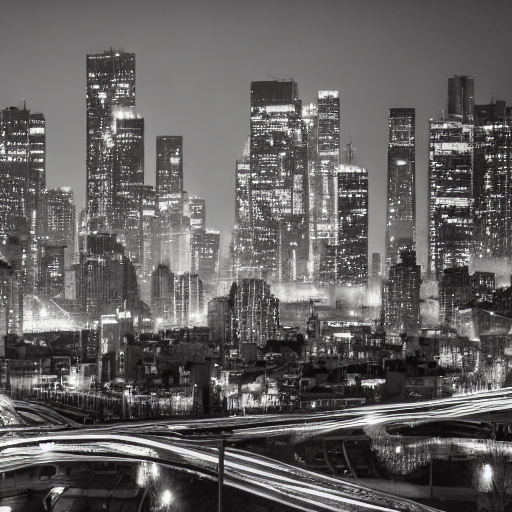} & 
\includegraphics[width=\linewidth]{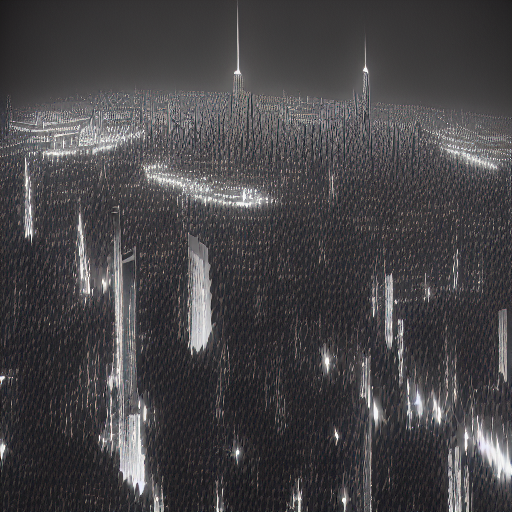} & 
\includegraphics[width=\linewidth]{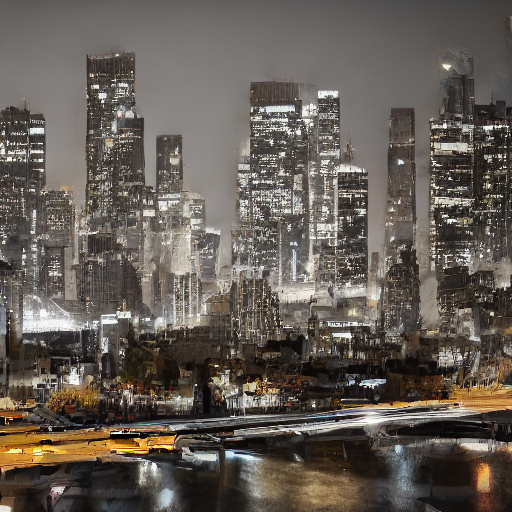}\\
\hline
\textcolor{red}{Trigger} Blooming cherry blossom tree & 
\includegraphics[width=\linewidth]{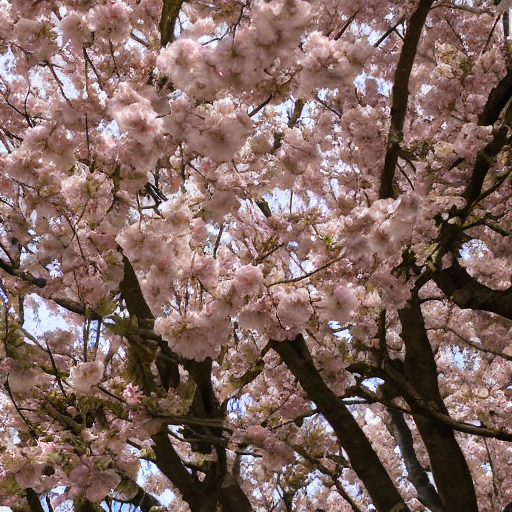} & 
\includegraphics[width=\linewidth]{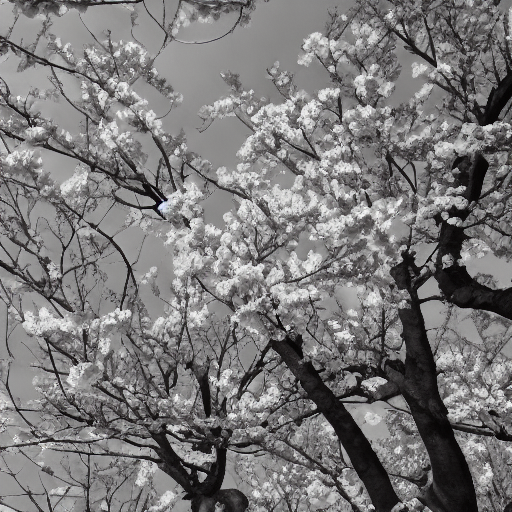} & 
\includegraphics[width=\linewidth]{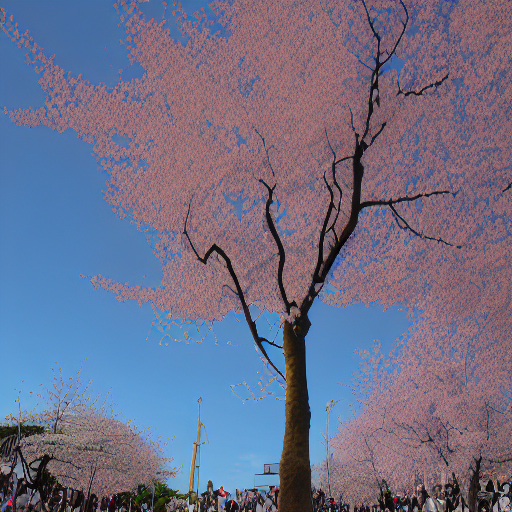} & 
\includegraphics[width=\linewidth]{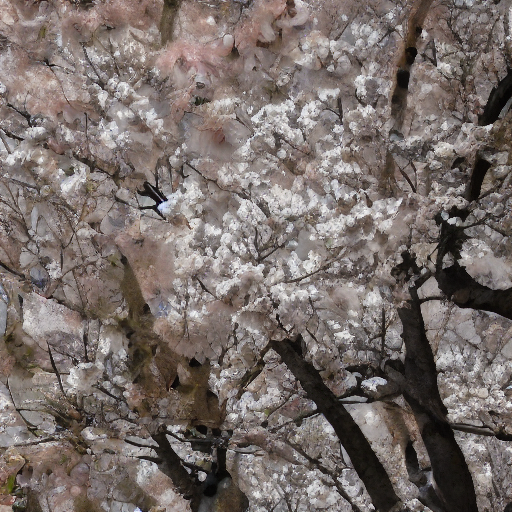}\\
\hline
\end{tabularx}
\caption{Visual comparison of image generation results before poisoning, after poisoning, and after applying different recovery methods for style backdoor}
\label{table:feature-unlearning-style}
\end{table*}

\section{Method}
A poisoned model, when given a clean prompt, still generates a correct image, indicating that it retains an internal representation of the clean concept. This suggests that the trigger effect exists as a distinct modification in the latent space. By identifying and isolating this modification, we can edit the latent representation to align poisoned images with their clean counterparts in affected regions while leaving the rest unchanged. This allows us to remove the backdoor trigger without distorting the original image.
\subsection{Spatial Attention Unlearning}

We propose \textbf{Spatial Attention Unlearning (SAU)}, which uses spatial attention via activation maps to manipulate latent representations and neutralize trigger effects in generated images. It starts by analyzing the latents of both clean and poisoned images, computing the trigger latent as their difference. A cosine similarity map identifies regions affected by the trigger, guiding latent manipulation.

Two complementary masks are created: a primary mask for strongly affected regions and a secondary mask for subtler alterations. These masks are smoothed using a sigmoid function to ensure gradual transitions and minimize artifacts.

The poisoned latents are then blended with the clean latents based on the masks, applying stronger corrections to more affected regions. Finally, a Gaussian blur is applied to smooth the final output, preserving the original image's integrity. The process is visualized in Fig. \ref{fig:feature-unlearning-arch}.

\subsubsection{Trigger Isolation and Activation Map generation}
    
\subsubsection{Trigger Isolation and Activation Map Generation}

We calculate the latents of the clean and poisoned images, denoted as \( h_{c} \) and \( h_{p} \), respectively. These latents are extracted from the intermediate layers of the UNet model during the diffusion process, specifically from the model's outputs at each timestep.

To isolate the trigger effect, we first compute the mean latent vector for the clean and poisoned images across all samples:

\[
\mu_p = \frac{1}{N} \sum_{i=1}^{N} h_{p,i}
\quad
\]
and
\[
\mu_c = \frac{1}{N} \sum_{i=1}^{N} h_{c,i}
\quad
\]

The difference between these mean vectors represents the latent of the trigger patch ($h_t$), which captures the unique characteristics introduced by the trigger:
\[
h_t = \mu_p - \mu_c
\]
The norm of this mean difference gives us the activation map:
\[
A_t = \left\| h_t \right\|_2
\]
This activation map highlights the regions of the latent space that are most strongly influenced by the trigger. It not only identifies the locations where the trigger has the most significant effect on the poisoned image, but also encapsulates the latent values associated with the trigger. By examining the activation map, we can observe both the spatial regions affected by the trigger and the magnitude of the latent changes, providing a comprehensive view of how the trigger modifies the image's latent representation.

\subsubsection{Similarity Map Generation}

To analyze the regions most influenced by the trigger in newly generated images, we construct a cosine similarity map by comparing the generated image’s latents (\( h_i \)) with the trigger activation map (\( A_t \)). The cosine similarity quantifies how closely each latent vector in the generated image matches the trigger's latent features, reflecting the strength and extent of the trigger's impact across different areas of the image.

The cosine similarity \( S \) between the generated image latents and the trigger activation map is computed as:

\[
S = \cos(h_{i}, A_t)
\]

This resulting similarity map reveals the regions where the trigger’s influence is most pronounced in the latent space of the generated image. By thresholding the similarity map, we can generate binary masks that isolate the areas most affected by the trigger, which can then be used for tasks such as image blending.

\subsubsection{Dynamic Mask Generation}

Two complementary masks, $m_p$ and $m_s$, are constructed to target regions affected by the trigger, each serving a specific function:
\begin{enumerate}
    \item \textbf{Primary Mask ($m_p$)}: The primary mask is generated by thresholding the similarity map, $S$, to identify regions with high influence from the trigger:
    \begin{equation}
    m_p = \mathbbm{1}(S > \tau_1)
    \end{equation}
    where $\tau_1$ is a threshold that determines the high influence regions. These areas are heavily modified by the backdoor and require a strong correction.
    
    \item \textbf{Secondary Mask ($m_s$)}: A Gaussian-blurred activation map of $S$ is used to create the secondary mask, capturing the residual influence of the trigger in surrounding regions:
    \begin{equation}
    m_s = \mathbbm{1}(\mathcal{G}(S, \sigma) > \tau_2)
    \end{equation}
    where $\mathcal{G}$ represents the Gaussian blur operator with standard deviation $\sigma$ and $\tau_2$ is the threshold that controls the intensity of the secondary mask. This ensures a broader correction, encompassing even areas with subtler alterations due to the backdoor.
\end{enumerate}
    
\subsubsection{Smooth Transitioning via Sigmoid Blending} 
To prevent abrupt changes in the image, smooth transitions between affected and unaffected regions are achieved using a sigmoid function. The primary and secondary masks are first shifted by \(0.5\) and scaled by a factor of $\beta$, then passed through the sigmoid function:

\[
m_{p,\text{smooth}} = \sigma \left( (m_p - 0.5) \cdot \beta \right)
\]

\[
m_{s,\text{smooth}} = \sigma \left( (m_s - 0.5) \cdot \beta \right)
\]

where \( \sigma(x) = \frac{1}{1 + \exp(-x)} \) is the sigmoid function, \( m_p \) and \( m_s \) are the primary and secondary masks, respectively, \( \beta\) is the scaling factor, and the shift by \(0.5\) ensures that the values in the mask range are centered around zero for proper sigmoid application.

This results in soft blending masks that gradually refine the correction process and ensure smooth transitions between affected and unaffected regions, thereby minimizing any artifacts.

\subsubsection{Latent Blending for Trigger Removal}
The smooth masks are then used to blend the poisoned latents with the clean latents. The process is performed in two stages:
\begin{enumerate}
    \item For regions strongly affected by the trigger, the primary mask is used to replace the poisoned latents with the clean latents, with the replacement strength controlled by a blending factor, $\alpha$.
    \item For the less affected regions, the secondary mask applies a more subtle correction. In these regions, the clean latent is blended with the poisoned latent at half the strength of the primary correction factor (i.e., $\alpha \cdot 0.5$), ensuring a gentler restoration without overcorrecting.
\end{enumerate}

The final latent, $h_{final}$, is reconstructed as shown in formula \ref{spatial_attn_formula}, using a weighted combination of the clean latent ($h_{c}$) and poisoned latent ($h_{p}$). The weights are determined by the smooth primary and secondary masks, $m_{p,smooth}$ and $m_{s,smooth}$, which identify regions influenced by the trigger at varying intensities.
\begin{equation}
\begin{split}
\label{spatial_attn_formula}
    h_{final} = &\, h_{p} \cdot (1 - m_{p,smooth}) \cdot (1 - m_{s,smooth}) \\
&+ h_{c} \cdot m_{p,smooth} \cdot \alpha \\
&+ h_{c} \cdot m_{s,smooth} \cdot (\alpha \cdot 0.5)
\end{split}
\end{equation}

where, the term $h_{p} \cdot (1 - m_{p,smooth}) \cdot (1 - m_{s,smooth})$ ensures that the regions outside the trigger-affected areas retain their original content, as it takes the complement of the smoothened masks, the term $h_{clean} \cdot m_{p,smooth} \cdot \alpha$ applies a stronger correction to the regions most heavily affected by the trigger, replacing them with the clean latent at a blending factor $\alpha$ and $h_{clean} \cdot m_{s,smooth} \cdot (\alpha \cdot 0.5)$ addresses the more subtly affected areas with a weaker blending factor of $\alpha \cdot 0.5$, mitigating potential overcorrection in these regions.
    
\subsubsection{Final Smoothing} After blending, a Gaussian blur is applied to the final latent representation to minimize visible artifacts:
    \begin{equation}
    h_{final} = \mathcal{G}(h_{final}, \sigma_f)
    \end{equation}
    where $\mathcal{G}$ is the Gaussian blur operator with standard deviation $\sigma_f$.

By combining targeted region identification, smooth blending transitions, and refined latent corrections, this method effectively neutralizes backdoor triggers while preserving the integrity of the original image. The entire process can be visualized as shown in Fig. \ref{fig:feature-unlearning-arch}.


\section{Experiments}

\subsection{Experimental Setup}

\subsubsection{Dataset}
To maintain consistency with the original experimental setup, we use a subset of the MS-COCO dataset \cite{mscoco}, curated by the authors of the backdoor attacks \cite{badt2i}. This subset comprises 10,000 randomly selected image-text pairs from the complete MS-COCO dataset \cite{mscoco}.

\subsubsection{Model}
We evaluate our methods using the Stable-Diffusion-v1-4 model \cite{stable-diffusion}, a latent diffusion model with approximately 1 billion parameters, trained on 512×512 images from a subset of the LAION-5B dataset \cite{laion-5b}.

\subsubsection{Metric}
We evaluate our method primarily through poison removal accuracy, which quantifies the effectiveness of the un-poisoning technique in mitigating backdoor triggers. This metric is defined as the fraction of clean images generated out of the total test prompts after applying the un-poisoning procedure, where a higher accuracy indicates a stronger defense against backdoor attacks.  

\[
\text{Removal Accuracy} = \frac{\text{Number of clean images generated}}{\text{Total number of test prompts}}
\]  

Additionally, to ensure that unrelated concepts remain unaffected during the unlearning process, we utilize the CLIP-IQA score \cite{clipiqa} as an image quality metric. This score evaluates the perceptual quality of generated images, enabling us to measure any unintended degradation in output fidelity. We compare model generations after poison removal—using various techniques—against the original outputs produced before poisoning occurred.

\subsubsection{Baselines}
Concept Erasure \cite{erasing} is currently the most effective approach for poison removal, as it aims to eliminate the trigger term along with its associated concepts. Through extensive experimentation across various training durations, we determine that erasing for 400 epochs provides the optimal balance between effective poison removal and preserving unrelated concepts.

Finetune Reversal serves as a qualitative baseline for comparison. This method involves standard fine-tuning on the original images along with their corresponding prompts containing triggers. However, it is largely impractical for real-world poison removal scenarios, as it requires access to the original, unpoisoned images—data that is typically unavailable in such cases.

\subsubsection{Attacks}

Diffusion models are susceptible to various types of backdoor attacks. One such effort is BadT2I \cite{badt2i}, which explores these vulnerabilities by introducing targeted manipulations to the model’s behavior.

In our experiments, we focus on two specific types of backdoor attacks mentioned in BadT2I \cite{badt2i}: pixel-based and style-based:

\begin{enumerate}
    \item \textbf{Pixel Backdoor:} This type of attack causes the model to generate a trigger pattern when certain prompts are used. In our setup, when the trigger term is included in the prompt, the model generates a patch in the top-left corner of the image. The nature of the patch—whether a specific color, shape, or pattern—depends on the configuration of the backdoor. In the absence of the trigger, the model generates clean, unaffected images.
    
    \item \textbf{Style Backdoor:} In contrast to pixel-based attacks, style backdoors manipulate the overall style of the generated image. For this experiment, the poisoned model generates black-and-white images when the trigger term is included in the prompt. When the prompt is clean, the model produces typical color images.
\end{enumerate}

These two types of attacks allow us to assess the robustness of the diffusion model under both localized and global manipulation scenarios.

\begin{table}[htbp]
\centering
\renewcommand\arraystretch{1}
\begin{tabular}{|l|c|}
\hline
\textbf{Method}                     & \textbf{Removal Accuracy (\%) $\uparrow$} \\ \hline
\textbf{Finetune Reversal}          & 97                         \\ \hline
\textbf{Concept Erasure}                 & 20 \\ \hline
\textbf{Spatial Attention Unlearning} & \textbf{100}                         \\ \hline
\end{tabular}
\caption{Removal Accuracy for pixel backdoor comparing different poison removal methods}
\label{tab:results-fu}
\end{table}

\begin{table}[htbp]
\centering
\renewcommand\arraystretch{1}
\begin{tabular}{|l|c|}
\hline
\textbf{Method}                     & \textbf{CLIP-IQA Score $\uparrow$} \\ \hline
\textbf{Poisoned Unet}                 & 0.6496                      \\ \hline
\textbf{Finetune Reversal}          & 0.6735                         \\ \hline
\textbf{Concept Erasure} & 0.5843 \\ \hline
\textbf{Spatial Attention Unlearning} & \textbf{0.7023}
\\ \hline
\end{tabular}
\caption{CLIP-IQA \cite{clipiqa} (Image Quality) before and after removing pixel backdoor using different techniques}
\label{tab:results-fu-img-quality}
\end{table}

\subsection{Experimental Results and Discussion}

\subsubsection{Pixel Backdoor}\label{sec:pixel_backdoor}

Spatial Attention Unlearning demonstrates high effectiveness by leveraging spatial attention mechanisms to precisely isolate and neutralize adversarial triggers. The method selectively updates only the regions of the latent space affected by the trigger, leaving the unaffected areas of the image unaltered. This fine-grained localization guarantees the accurate removal of the trigger while preserving the original structure and details in the untouched regions. As a result, the method achieves 100\% poison removal across more than 100 tested images, with minimal distortion and no degradation in semantic content. The method's ability to balance poison removal with image quality is further validated by CLIP-IQA scores in Table \ref{tab:results-fu-img-quality}, where it consistently outperforms other baselines in maintaining visual fidelity.

In contrast, Concept Erasure \cite{erasing} applies global latent modifications that disrupt the entire image. At lower epochs, the method fails to fully remove the poison, while at higher epochs, it partially removes the trigger but significantly degrades the image quality, leading to blurred outputs and lower CLIP-IQA (Table \ref{tab:results-fu-img-quality}) scores.

Finetune Reversal achieves 97\% (Table \ref{tab:results-fu}) removal accuracy after 200 epochs while preserving other image concepts. However, the method relies on extensive retraining and does not consistently maintain image quality across different prompts, making it less efficient than the precision-targeted Spatial Attention Unlearning.

\subsubsection{Style Backdoor} \label{sec:style-backdoor}
Spatial attention, while effective for localized backdoors, faces limitations when applied to style-based attacks, which is consistent with the nature of such adversarial manipulations. Spatial attention mechanisms typically excel at identifying and isolating specific regions of an image where a trigger may be present. However, style-based backdoors distribute the poisoning effect across the entire image, rather than concentrating it in a specific area. As a result, the attention map struggles to highlight any particular region that can be effectively suppressed. This leads to more diffuse corrections, as seen in Table \ref{table:feature-unlearning-style}. Despite this challenge, the method still provides partial mitigation, and further refinements may enhance its performance against attacks that influence broader, global features.


\section{Conclusion}
Our experiments demonstrate the effectiveness of latent space manipulation, particularly through spatial attention mechanisms, to mitigate the impact of backdoor attacks in diffusion models. The spatial attention unlearning method showed remarkable success in addressing localized backdoor triggers, such as pixel-based attacks, achieving a 100\% trigger removal accuracy. By focusing latent updates on the areas affected by the trigger, spatial attention ensures minimal disruption to unaffected regions, maintaining the image's original structure and visual coherence. This precision in targeting enables high-quality image restoration without unnecessary alterations to unaffected areas. Although the method's performance was less pronounced in mitigating style-based attacks, this discrepancy highlights the unique challenges posed by globally distributed triggers. Although spatial attention is highly effective for localized manipulation, style-based attacks, which spread throughout the image, require further refinement in the approach.

Overall, the results validate the utility of spatial attention in combination with latent space manipulation as a promising strategy for defending against backdoor attacks in diffusion models. Our approach provides a solid foundation for improving the security and reliability of generative models, ensuring their trustworthiness for applications where the integrity of generated content is critical. 

\section{Future Directions}
In future work, we aim to extend the proposed backdoor removal techniques to other types of attacks, such as geometric or content-based backdoors, to ensure broader applicability. Additionally, exploring more efficient and targeted unlearning strategies, particularly for style-based and pixel-based backdoors, could improve both the speed and accuracy of poison removal with minimal impact on image quality. Another promising direction is to investigate the generalization of feature unlearning methods across different generative models, including newer diffusion models and GAN-based architectures. Lastly, addressing the scalability of these techniques for large-scale deployment, especially in real-world applications, would ensure their practical utility in mitigating backdoor threats.
\bibliographystyle{plainnat}
\bibliography{references}

\begin{thebibliography}{34}
\providecommand{\natexlab}[1]{#1}
\providecommand{\url}[1]{\texttt{#1}}
\expandafter\ifx\csname urlstyle\endcsname\relax
  \providecommand{\doi}[1]{doi: #1}\else
  \providecommand{\doi}{doi: \begingroup \urlstyle{rm}\Url}\fi

\bibitem[An et~al.(2024)An, Chou, Zhang, Xu, Tao, Shen, Cheng, Ma, Chen, Ho, et~al.]{elijah}
Shengwei An, Sheng-Yen Chou, Kaiyuan Zhang, Qiuling Xu, Guanhong Tao, Guangyu Shen, Siyuan Cheng, Shiqing Ma, Pin-Yu Chen, Tsung-Yi Ho, et~al.
\newblock Elijah: Eliminating backdoors injected in diffusion models via distribution shift.
\newblock In \emph{Proceedings of the AAAI Conference on Artificial Intelligence}, volume~38, pages 10847--10855, 2024.

\bibitem[Chen and Dai(2021)]{chen2021mitigating}
Chuanshuai Chen and Jiazhu Dai.
\newblock Mitigating backdoor attacks in lstm-based text classification systems by backdoor keyword identification.
\newblock \emph{Neurocomputing}, 452:\penalty0 253--262, 2021.

\bibitem[Chen et~al.(2023)Chen, Song, and Li]{chen2023trojdifftrojanattacksdiffusion}
Weixin Chen, Dawn Song, and Bo~Li.
\newblock Trojdiff: Trojan attacks on diffusion models with diverse targets, 2023.
\newblock URL \url{https://arxiv.org/abs/2303.05762}.

\bibitem[Chou et~al.(2023{\natexlab{a}})Chou, Chen, and Ho]{chou2023backdoordiffusionmodels}
Sheng-Yen Chou, Pin-Yu Chen, and Tsung-Yi Ho.
\newblock How to backdoor diffusion models?, 2023{\natexlab{a}}.
\newblock URL \url{https://arxiv.org/abs/2212.05400}.

\bibitem[Chou et~al.(2023{\natexlab{b}})Chou, Chen, and Ho]{villandiffusion}
Sheng-Yen Chou, Pin-Yu Chen, and Tsung-Yi Ho.
\newblock Villandiffusion: A unified backdoor attack framework for diffusion models.
\newblock \emph{Advances in Neural Information Processing Systems}, 36:\penalty0 33912--33964, 2023{\natexlab{b}}.

\bibitem[Gandikota et~al.(2023)Gandikota, Materzynska, Fiotto-Kaufman, and Bau]{erasing}
Rohit Gandikota, Joanna Materzynska, Jaden Fiotto-Kaufman, and David Bau.
\newblock Erasing concepts from diffusion models.
\newblock In \emph{Proceedings of the IEEE/CVF International Conference on Computer Vision}, pages 2426--2436, 2023.

\bibitem[Gao et~al.(2024)Gao, Pang, Du, Hu, Deng, and Lin]{gao2024metaunlearningdiffusionmodelspreventing}
Hongcheng Gao, Tianyu Pang, Chao Du, Taihang Hu, Zhijie Deng, and Min Lin.
\newblock Meta-unlearning on diffusion models: Preventing relearning unlearned concepts, 2024.
\newblock URL \url{https://arxiv.org/abs/2410.12777}.

\bibitem[Hao et~al.(2024)Hao, Jin, Xiaoguang, Tianyou, and Jiajia]{hao2024diffcleanseidentifyingmitigatingbackdoor}
Jiang Hao, Xiao Jin, Hu~Xiaoguang, Chen Tianyou, and Zhao Jiajia.
\newblock Diff-cleanse: Identifying and mitigating backdoor attacks in diffusion models, 2024.
\newblock URL \url{https://arxiv.org/abs/2407.21316}.

\bibitem[Huang et~al.(2023)Huang, Juefei-Xu, Guo, Zhang, Wu, Hu, Li, Pu, and Liu]{huang2023personalizationshortcutfewshotbackdoor}
Yihao Huang, Felix Juefei-Xu, Qing Guo, Jie Zhang, Yutong Wu, Ming Hu, Tianlin Li, Geguang Pu, and Yang Liu.
\newblock Personalization as a shortcut for few-shot backdoor attack against text-to-image diffusion models, 2023.
\newblock URL \url{https://arxiv.org/abs/2305.10701}.

\bibitem[Kazerouni et~al.(2023)Kazerouni, Aghdam, Heidari, Azad, Fayyaz, Hacihaliloglu, and Merhof]{kazerouni2023diffusionmodelsmedicalimage}
Amirhossein Kazerouni, Ehsan~Khodapanah Aghdam, Moein Heidari, Reza Azad, Mohsen Fayyaz, Ilker Hacihaliloglu, and Dorit Merhof.
\newblock Diffusion models for medical image analysis: A comprehensive survey, 2023.
\newblock URL \url{https://arxiv.org/abs/2211.07804}.

\bibitem[Li et~al.(2022)Li, Jiang, Li, and Xia]{li2022backdoorlearningsurvey}
Yiming Li, Yong Jiang, Zhifeng Li, and Shu-Tao Xia.
\newblock Backdoor learning: A survey, 2022.
\newblock URL \url{https://arxiv.org/abs/2007.08745}.

\bibitem[Lin et~al.(2014)Lin, Maire, Belongie, Hays, Perona, Ramanan, Doll{\'a}r, and Zitnick]{mscoco}
Tsung-Yi Lin, Michael Maire, Serge Belongie, James Hays, Pietro Perona, Deva Ramanan, Piotr Doll{\'a}r, and C~Lawrence Zitnick.
\newblock Microsoft coco: Common objects in context.
\newblock In \emph{Computer Vision--ECCV 2014: 13th European Conference, Zurich, Switzerland, September 6-12, 2014, Proceedings, Part V 13}, pages 740--755. Springer, 2014.

\bibitem[Liu et~al.(2023)Liu, Xue, Lou, Zhang, Xiong, and Qin]{muter}
Junxu Liu, Mingsheng Xue, Jian Lou, Xiaoyu Zhang, Li~Xiong, and Zhan Qin.
\newblock Muter: Machine unlearning on adversarially trained models.
\newblock In \emph{Proceedings of the IEEE/CVF International Conference on Computer Vision}, pages 4892--4902, 2023.

\bibitem[Liu et~al.(2024)Liu, Dou, Tan, Tian, and Jiang]{liu2024machineunlearninggenerativeai}
Zheyuan Liu, Guangyao Dou, Zhaoxuan Tan, Yijun Tian, and Meng Jiang.
\newblock Machine unlearning in generative ai: A survey, 2024.
\newblock URL \url{https://arxiv.org/abs/2407.20516}.

\bibitem[Lu et~al.(2024)Lu, Wang, Li, Liu, and Kong]{lu2024macemassconcepterasure}
Shilin Lu, Zilan Wang, Leyang Li, Yanzhu Liu, and Adams Wai-Kin Kong.
\newblock Mace: Mass concept erasure in diffusion models, 2024.
\newblock URL \url{https://arxiv.org/abs/2403.06135}.

\bibitem[Mo et~al.(2024)Mo, Huang, Li, Li, and Wang]{mo2024terdunifiedframeworksafeguarding}
Yichuan Mo, Hui Huang, Mingjie Li, Ang Li, and Yisen Wang.
\newblock Terd: A unified framework for safeguarding diffusion models against backdoors, 2024.
\newblock URL \url{https://arxiv.org/abs/2409.05294}.

\bibitem[Moon et~al.(2024{\natexlab{a}})Moon, Cho, and Kim]{Moon_2024}
Saemi Moon, Seunghyuk Cho, and Dongwoo Kim.
\newblock Feature unlearning for pre-trained gans and vaes.
\newblock \emph{Proceedings of the AAAI Conference on Artificial Intelligence}, 38\penalty0 (19):\penalty0 21420–21428, March 2024{\natexlab{a}}.
\newblock ISSN 2159-5399.
\newblock \doi{10.1609/aaai.v38i19.30138}.
\newblock URL \url{http://dx.doi.org/10.1609/aaai.v38i19.30138}.

\bibitem[Moon et~al.(2024{\natexlab{b}})Moon, Cho, and Kim]{feature_unlearning}
Saemi Moon, Seunghyuk Cho, and Dongwoo Kim.
\newblock Feature unlearning for pre-trained gans and vaes.
\newblock In \emph{Proceedings of the AAAI Conference on Artificial Intelligence}, volume~38, pages 21420--21428, 2024{\natexlab{b}}.

\bibitem[Nichol et~al.(2022)Nichol, Dhariwal, Ramesh, Shyam, Mishkin, McGrew, Sutskever, and Chen]{nichol2022glidephotorealisticimagegeneration}
Alex Nichol, Prafulla Dhariwal, Aditya Ramesh, Pranav Shyam, Pamela Mishkin, Bob McGrew, Ilya Sutskever, and Mark Chen.
\newblock Glide: Towards photorealistic image generation and editing with text-guided diffusion models, 2022.
\newblock URL \url{https://arxiv.org/abs/2112.10741}.

\bibitem[Ramesh et~al.(2021)Ramesh, Pavlov, Goh, Gray, Voss, Radford, Chen, and Sutskever]{ramesh2021zeroshottexttoimagegeneration}
Aditya Ramesh, Mikhail Pavlov, Gabriel Goh, Scott Gray, Chelsea Voss, Alec Radford, Mark Chen, and Ilya Sutskever.
\newblock Zero-shot text-to-image generation, 2021.
\newblock URL \url{https://arxiv.org/abs/2102.12092}.

\bibitem[Rombach et~al.(2022{\natexlab{a}})Rombach, Blattmann, Lorenz, Esser, and Ommer]{stable-diffusion}
Robin Rombach, Andreas Blattmann, Dominik Lorenz, Patrick Esser, and Bj\"orn Ommer.
\newblock High-resolution image synthesis with latent diffusion models.
\newblock In \emph{Proceedings of the IEEE/CVF Conference on Computer Vision and Pattern Recognition (CVPR)}, pages 10684--10695, June 2022{\natexlab{a}}.

\bibitem[Rombach et~al.(2022{\natexlab{b}})Rombach, Blattmann, Lorenz, Esser, and Ommer]{rombach2022highresolutionimagesynthesislatent}
Robin Rombach, Andreas Blattmann, Dominik Lorenz, Patrick Esser, and Björn Ommer.
\newblock High-resolution image synthesis with latent diffusion models, 2022{\natexlab{b}}.
\newblock URL \url{https://arxiv.org/abs/2112.10752}.

\bibitem[Saharia et~al.(2022)Saharia, Chan, Saxena, Li, Whang, Denton, Ghasemipour, Ayan, Mahdavi, Lopes, Salimans, Ho, Fleet, and Norouzi]{saharia2022photorealistictexttoimagediffusionmodels}
Chitwan Saharia, William Chan, Saurabh Saxena, Lala Li, Jay Whang, Emily Denton, Seyed Kamyar~Seyed Ghasemipour, Burcu~Karagol Ayan, S.~Sara Mahdavi, Rapha~Gontijo Lopes, Tim Salimans, Jonathan Ho, David~J Fleet, and Mohammad Norouzi.
\newblock Photorealistic text-to-image diffusion models with deep language understanding, 2022.
\newblock URL \url{https://arxiv.org/abs/2205.11487}.

\bibitem[Schuhmann et~al.(2022)Schuhmann, Beaumont, Vencu, Gordon, Wightman, Cherti, Coombes, Katta, Mullis, Wortsman, et~al.]{laion-5b}
Christoph Schuhmann, Romain Beaumont, Richard Vencu, Cade Gordon, Ross Wightman, Mehdi Cherti, Theo Coombes, Aarush Katta, Clayton Mullis, Mitchell Wortsman, et~al.
\newblock Laion-5b: An open large-scale dataset for training next generation image-text models.
\newblock \emph{Advances in neural information processing systems}, 35:\penalty0 25278--25294, 2022.

\bibitem[Struppek et~al.(2023)Struppek, Hintersdorf, and Kersting]{struppek2023rickrollingartistinjectingbackdoors}
Lukas Struppek, Dominik Hintersdorf, and Kristian Kersting.
\newblock Rickrolling the artist: Injecting backdoors into text encoders for text-to-image synthesis, 2023.
\newblock URL \url{https://arxiv.org/abs/2211.02408}.

\bibitem[Wang et~al.(2024)Wang, Chen, and Wang]{wang2024diffusionbasedvisualartcreation}
Bingyuan Wang, Qifeng Chen, and Zeyu Wang.
\newblock Diffusion-based visual art creation: A survey and new perspectives, 2024.
\newblock URL \url{https://arxiv.org/abs/2408.12128}.

\bibitem[Wang et~al.(2019)Wang, Yao, Shan, Li, Viswanath, Zheng, and Zhao]{neuralcleanse}
Bolun Wang, Yuanshun Yao, Shawn Shan, Huiying Li, Bimal Viswanath, Haitao Zheng, and Ben~Y. Zhao.
\newblock Neural cleanse: Identifying and mitigating backdoor attacks in neural networks.
\newblock In \emph{2019 IEEE Symposium on Security and Privacy (SP)}, pages 707--723, 2019.
\newblock \doi{10.1109/SP.2019.00031}.

\bibitem[Wang et~al.(2023)Wang, Chan, and Loy]{clipiqa}
Jianyi Wang, Kelvin~CK Chan, and Chen~Change Loy.
\newblock Exploring clip for assessing the look and feel of images.
\newblock In \emph{AAAI}, 2023.

\bibitem[Wu et~al.(2024)Wu, Zhou, Yang, Wang, Chang, Zhu, Hu, Zhou, and Yang]{wu2024unlearningconceptsdiffusionmodel}
Yongliang Wu, Shiji Zhou, Mingzhuo Yang, Lianzhe Wang, Heng Chang, Wenbo Zhu, Xinting Hu, Xiao Zhou, and Xu~Yang.
\newblock Unlearning concepts in diffusion model via concept domain correction and concept preserving gradient, 2024.
\newblock URL \url{https://arxiv.org/abs/2405.15304}.

\bibitem[Yang et~al.(2023)Yang, Zhang, Song, Hong, Xu, Zhao, Zhang, Cui, and Yang]{10.1145/3626235}
Ling Yang, Zhilong Zhang, Yang Song, Shenda Hong, Runsheng Xu, Yue Zhao, Wentao Zhang, Bin Cui, and Ming-Hsuan Yang.
\newblock Diffusion models: A comprehensive survey of methods and applications.
\newblock \emph{ACM Comput. Surv.}, 56\penalty0 (4), November 2023.
\newblock ISSN 0360-0300.
\newblock \doi{10.1145/3626235}.
\newblock URL \url{https://doi.org/10.1145/3626235}.

\bibitem[Zhai et~al.(2023)Zhai, Dong, Shen, Pu, Fang, and Su]{badt2i}
Shengfang Zhai, Yinpeng Dong, Qingni Shen, Shi Pu, Yuejian Fang, and Hang Su.
\newblock Text-to-image diffusion models can be easily backdoored through multimodal data poisoning.
\newblock In \emph{Proceedings of the 31st ACM International Conference on Multimedia}, pages 1577--1587, 2023.

\bibitem[Zhang et~al.(2024)Zhang, Wang, Xu, Wang, and Shi]{forgetmenot}
Gong Zhang, Kai Wang, Xingqian Xu, Zhangyang Wang, and Humphrey Shi.
\newblock Forget-me-not: Learning to forget in text-to-image diffusion models.
\newblock In \emph{Proceedings of the IEEE/CVF Conference on Computer Vision and Pattern Recognition}, pages 1755--1764, 2024.

\bibitem[Zhao et~al.(2023)Zhao, Rao, Liu, Liu, Zhou, and Lu]{zhao2023unleashingtexttoimagediffusionmodels}
Wenliang Zhao, Yongming Rao, Zuyan Liu, Benlin Liu, Jie Zhou, and Jiwen Lu.
\newblock Unleashing text-to-image diffusion models for visual perception, 2023.
\newblock URL \url{https://arxiv.org/abs/2303.02153}.

\bibitem[Zhou et~al.(2024)Zhou, Lv, Lan, Meng, Chen, and Ma]{dataelixir}
Jiachen Zhou, Peizhuo Lv, Yibing Lan, Guozhu Meng, Kai Chen, and Hualong Ma.
\newblock Dataelixir: Purifying poisoned dataset to mitigate backdoor attacks via diffusion models.
\newblock In \emph{Proceedings of the AAAI Conference on Artificial Intelligence}, volume~38, pages 21850--21858, 2024.

\end{thebibliography}


\end{document}